\documentclass[aps,twocolumn,showpacs,amsmath]{revtex4}
\usepackage{graphicx}
\begin{document}
\setlength{\arraycolsep}{2pt}
\title{Entanglement within the Quantum Trajectory Description of Open Quantum Systems}
\author{Hyunchul Nha and H. J. Carmichael}
\affiliation{Department of Physics, University of Auckland, Private Bag 92019,
Auckland, New Zealand} 

\begin{abstract}
The degree of entanglement in an open quantum system varies according to how
information in the environment is read. A measure of this contextual entanglement
is introduced based on quantum trajectory unravelings of the open system dynamics.
It is used to characterize the entanglement in a driven quantum system of dimension
$2\times\infty$ where the entanglement is induced by the environmental interaction.
A detailed  mechanism for the environment-induced entanglement is given. 
\end{abstract}
\maketitle
\narrowtext

Entanglement is a remarkable feature of quantum mechanics that has attracted
much attention in recent years for its potential use as a resource in quantum
information processing \cite{Nielsen00}. In practical situations a quantum
system inevitably couples to its environment, so that the state of the system
becomes mixed. Much effort has therefore been invested in finding a reasonable
measure of entanglement for mixed states. Specific proposals include the
entanglement of formation \cite{Bennett96,Wootters98} and distillation
\cite{Bennett96}, and the relative entropy of entanglement \cite{Vedral97}.
These measures are not practicably computable in general, however, and are
readily accessible only for low-dimensional systems \cite{Wootters98,Vedral97}
and certain symmetric states \cite{Terhal00}. Recently, a variational method
for calculating the entanglement of formation (EOF) in general bipartite states
was developed \cite{Audenaert01}. For bipartite mixed states, a computable
measure of entanglement based on the negativity of the partial transpose has
also been proposed \cite{Vidal02}. In the case of continuous variable systems,
the EOF of symmetric two-mode Gaussian states was recently obtained
\cite{Giedke03}.

In this Letter we consider mixed state entanglement from a new perspective.
For pure states, the accepted measure of entanglement is settled. Considering
a bipartite state $\rho_i=|\psi_i\rangle\langle\psi_i|$ (subsystems $A$ and
$B$), it is calculated as the von Neumann entropy
\begin{equation}
E(\rho_i^B)=-{\rm tr}_B\!\left[\rho_i^B{\rm log_2}(\rho_i^B)\right],
\label{eqn:pure_entanglement}
\end{equation}
where $\rho_i^B={\rm tr}_A(|\psi_i\rangle\langle\psi_i|)$ \cite{Bennett96}.
It is well known that any mixed state $\rho$ may be decomposed into an
ensemble of pure states in an infinity of ways: $\rho=\Sigma_ip_i
|\psi_i\rangle\langle\psi_i|$, where $p_i\ge0$ and $\Sigma_ip_i=1$. Given a
decomposition, the entanglement may be quantified by the ensemble average
$E=\Sigma_ip_iE(\rho_i^B)$. The perceived difficulty here is with the
arbitrariness of the decomposition. Measures of mixed state entanglement
differ in their strategy for replacing this arbitrariness by something
specific. The EOF, for example, is defined as the minimum over all
decompositions; it intends to quantify the resources required to produce
the state $\rho$ under a specified quantum communication protocol
\cite{Bennett96,Wootters98}. The work reported in
this Letter takes a different view. We observe that entangled states are
pervasive in the quantum mechanics of composite systems, often arising in
situations that have no immediate quantum information connection. With the
aim of understanding entanglement in this broader sense, we propose to
explore, rather than discard, the multiplicity of mixed state
decompositions. The ambiguity arising in the multiplicity is viewed as a
signature of complementarity, i.e., an expected feature that reflects the
very essence of a quantum mechanical description.

We consider entanglement in bipartite open systems, specifically, where
we resolve the arbitrariness in the decomposition of $\rho$ by limiting
attention to physically relevant decompositions---those for which $p_i$
is the probability of a classical record comprising information read from
the environment. This information may be read in many ways.
For each, there is a different decomposition of $\rho$ and hence a different
degree of entanglement. Quantum trajectory theory \cite{Carmichael93}
provides a natural measure of this entanglement in the form
\begin{equation}
E_U=\overline{E(\rho_{U;{\rm REC}}^B)},
\label{eqn:unravelling_entanglement}
\end{equation}
with $\rho_{U;{\rm REC}}^B={\rm tr}_A(|\psi_{U;{\rm REC}}\rangle
\langle\psi_{U;{\rm REC}}|)$, where $U$ labels a particular unraveling of
the open system dynamics (reading of the environment) and ${\rm REC}$ denotes
a particular record; the overbar in Eq.~(\ref{eqn:unravelling_entanglement})
denotes an average over records. The reading of the environment is continuous
in time, and to preserve the purity of the state $|\psi_{U;{\rm REC}}\rangle$,
$100\%$ efficient (all scattered particles are ultimately detected). $E_U$ has a
particular significance for each physically realizable unraveling. 

We explore these ideas in an example, where we characterize the entanglement
in a composite system consisting of ($A$) an optical cavity mode (harmonic
oscillator), resonantly coupled to ($B$) a two-state atom (single qubit).
The atom is driven by a resonant external field and both subsystems are
coupled to Markov reservoirs to account for the scattering of light into the
vacuum of the electromagnetic field. The Hamiltonian is
\begin{eqnarray}
H=H_{AB}+H_{\rm ext}^B+H_{\rm res},
\end{eqnarray}
with
\begin{eqnarray}
H_{AB}&=&\hbar\omega(\hat a^{\dag}\hat a+\hat b^{\dag}\hat b)
+i\hbar g(\hat a^{\dag}\hat b-\hat b^{\dag}\hat a),\nonumber\\
H_{\rm ext}^B&=&i\hbar\Omega(\hat b^{\dag}e^{-i\omega t}-\hat b
e^{i\omega t}),\nonumber\\
H_{\rm res}&=&H_{R_A}+H_{R_B}+H_{AR_A}+H_{BR_B},
\end{eqnarray}
 where
$\hat a$ and $\hat a^\dagger$ are cavity mode annihilation and creation
operators ($[\hat a,\hat a^\dagger]=1$), and $\hat b$ and $\hat
b^{\dag}$ are lowering and raising operators for the atom ($[\hat b,
\hat b^\dagger]_+=1$). The external field has amplitude $\Omega$, and $A$
and $B$ couple to reservoirs $R_A$ and $R_B$, respectively, via interactions
$H_{AR_A}$ and $H_{BR_B}$. The evolution of the reduced density
operator $\tilde\rho$ in the interaction picture is governed by the master
equation 
\begin{eqnarray}
\frac{d\tilde\rho}{d\bar t}=\frac{1}{i\hbar}[\tilde H_{AB}+
\tilde H_{\rm ext}^B,\tilde\rho]+\left({\cal L}_A+{\cal L}_B\right)\tilde\rho,
\label{eqn:master1}
\end{eqnarray}
where $\tilde H_{AB}=i\hbar(\hat a^{\dag}\hat b-\hat b^{\dag}\hat a)$,
$\tilde H_{\rm ext}^B=i\hbar\bar\Omega(\hat b^{\dag}-\hat b)$, with
$\bar\Omega\equiv\Omega/g$, $\bar t\equiv gt$, and
\begin{eqnarray}
{\cal L}_{A,B}=\bar \Gamma_{a,b}(2\hat o\cdot \hat o^{\dag}-\hat o^{\dag}
\hat o\cdot-\cdot \hat o^{\dag}\hat o),
\end{eqnarray}
where $(\hat o=\hat a,\hat b)$ and 
the damping constants $\bar\Gamma_{a,b}\equiv\Gamma_{a,b}/g$ determine
the reservoir interaction strengths. 

The dimension of $A\otimes B$ is $2\times\infty$; hence the results
of Refs.~\cite{Vedral97,Terhal00,Audenaert01,Giedke03} do not apply. The
system exhibits the additional interesting feature of {\it environment-assisted\/}
entanglement---i.e., an interaction with the environment is required to
generate entanglement. Specifically, the degree of entanglement increases
with the damping of the cavity mode $A$, passes through a peak, and then
decreases again. The behavior illustrates something of the subtlety associated
with entanglement generation in open systems; the environment is not simply
a source of decoherence. Squeezing in the considered system shows a similar
counterintuitive dependence on the cavity damping \cite{Nha03}. Entanglement
generation through an environmental interaction has also been reported for
two cavity modes coupled to an incoherently driven
atom \cite{Plenio02} and for two atoms coupled to a common bath \cite{Benatti03}.

We first uncover the origin of the environment-assisted entanglement.
We note that $\tilde H_{AB}+\tilde H_{\rm ext}^B$ has an eigenstate
$|\alpha\rangle|g\rangle$, $\alpha=\bar\Omega$, where $|\alpha\rangle$ is
a coherent state of  $A$ and $|g\rangle$ denotes the ground state of $B$.
The interpretation is that since the atom interacts with the sum of fields
$\hat a-\bar\Omega$, it is stable in its ground state if the summed field
amplitude is zero. In the presence of damping $\bar\Gamma_b\neq0$, but with
$\bar\Gamma_a=0$, this eigenstate becomes the system steady state (a {\it dark
state\/}) for all $\bar\Gamma_b$ and $\bar\Omega$ \cite{Alsing92}. Thus, if
there is no interaction between subsystem $A$ and its environment the system
steady state is the product state $|\bar\Omega\rangle|g\rangle$, and the
entanglement $E_U$ is zero; effectively the coupling between subsystems
is turned off. The environmental interaction $\bar\Gamma_a$ destabilizes the
dark state, restores the coupling, and generates entanglement.

To describe the situation for $\bar\Gamma_a\neq0$ it is convenient to introduce
the displaced state $\tilde\rho^\prime$, with $\tilde{\rho}=\hat D^\dagger
(\bar\Omega)\tilde\rho^\prime\hat D(\bar\Omega)$, where $\hat D(\bar\Omega)
\equiv\exp[\bar\Omega(\hat a^\dag-\hat a)]$. Since $\hat D(\bar\Omega)$ is
a local unitary operator, it does not change the degree of entanglement. The
displacement moves the driving from $B$ to $A$ \cite{Alsing91}. In place of
Eq.~(\ref{eqn:master1}), we obtain
\begin{eqnarray}
\frac{d\tilde{\rho}^\prime}{d\bar t}=\frac{1}{i\hbar}[\tilde H_{AB}
+\tilde H_{\rm ext}^A,\tilde{\rho}^\prime]+\left({\cal L}_A+{\cal L}_B
\right)\tilde{\rho}^\prime,
\label{eqn:master2}
\end{eqnarray}
where $\tilde H_{\rm ext}^A=i\hbar\bar\Gamma_a\bar\Omega(\hat a-
\hat a^{\dag})$. Note that $\bar\Gamma_a$ appears in two places:
in the driving field interaction $\tilde H_{\rm ext}^A$, where it governs
the described creation of entanglement, and in the usual damping term
${\cal L}_A\tilde\rho^\prime$. The interplay of the two, along with the
interaction $\hat H_{AB}$, determines the behavior of the entanglement as a
function of $\bar\Gamma_a$. In particular, in the limit $\bar\Gamma_a\to\infty$,
the steady state of the cavity field in the displaced frame balances $\tilde
H_{\rm ext}^A$ with ${\cal L}_A$. This yields a coherent state of amplitude
$\alpha^\prime=-\bar\Omega$, the vacuum state in the original frame (see 
Fig.~\ref{fig:fig2}). Thus, in the large $\bar\Gamma_a$ limit, the steady
state approaches $\tilde\rho=\tilde\rho_A\tilde\rho_B$, where $\tilde\rho_A
=|0\rangle\langle0|$ and $\tilde\rho_B$ is determined by the balancing of
$\tilde H_{\rm ext}^B$ with ${\cal L}_B$, i.e., $\tilde\rho_B$ is the steady
state of resonance fluorescence. In summary, for $\bar\Gamma_a=0$ or $\infty$ 
the entanglement is zero.

\begin{figure}
\includegraphics*[width=3.2in,keepaspectratio=true]{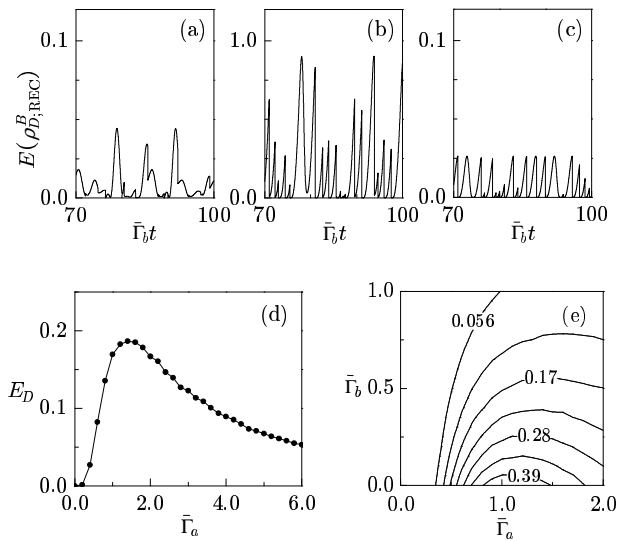}
\caption{Entanglement in direct detection, for $\bar\Omega=1$: typical
temporal behavior of the von Neumann entropy for $\bar\Gamma_b=0.5$ and
$\bar\Gamma_a=0.3$ (a), $2.0$ (b), and $20.0$ (c); (d) entanglement
[Eq.~(\ref{eqn:unravelling_entanglement})] as a function of $\bar\Gamma_a$
(for $\bar\Gamma_b=0.5$); (e) entanglement as a function of $\bar\Gamma_a$ and
$\bar\Gamma_b$.}
\label{fig:fig1}
\end{figure}

Consider now the quantification of entanglement in between these limits. Figure
\ref{fig:fig1} illustrates the behavior with the unraveling of the density operator
based upon direct detection of the scattered photons (unraveling $U=D$),
those scattered through the cavity mirrors via the coupling of $B$ to $A$ and
those scattered directly by the atom $B$. Frames (a)-(c) illustrate the time
dependence of the von Neumann entropy, where the entanglement is conditioned on
a particular record of photon counts. Frames (d) and (e) show results after
taking the average over records (a time average is permitted since trajectories
are ergodic \cite{Kummerer03,Cresser01}). In the small and large damping limits,
the entanglement is calculated to be
\begin{equation}
E_D=-\lambda{\rm log}_2\lambda-(1-\lambda){\rm log}_2(1-\lambda),
\label{eqn:entropy}
\end{equation}
with $\lambda\approx\bar\Omega^4\bar\Gamma_a^6$ and $\lambda\approx
(2/\bar\Gamma_a)^2(4\bar\Omega/\bar\Gamma_b)^4$, respectively. The peak
in between these limits is shown in Figs.~\ref{fig:fig1}(d) and (e), and
is to be compared with the monotonic decrease of $E_D$ with increasing
$\bar\Gamma_b$. (Increasing $\bar\Gamma_b$ increases the rate of
interruptions, due to the detection of a photon in the environment, that
return the atom to the ground state; each interruption destroys any
entanglement created since the previous photon detection.)

We turn now to the variation in the degree of entanglement for different
unravelings ($U$) of $\rho$.
For simplicity, we set $\bar\Gamma_b=0$ and
focus attention on the behavior of the entanglement as a function of
$\bar\Gamma_a$. We begin by presenting an analytical approximation, which
clarifies the behavior in Fig.~\ref{fig:fig1}(d) and serves as an
introduction to the unraveling-dependence of the entanglement.

\begin{figure}[b]
\includegraphics*[width=3.2in,keepaspectratio=true]{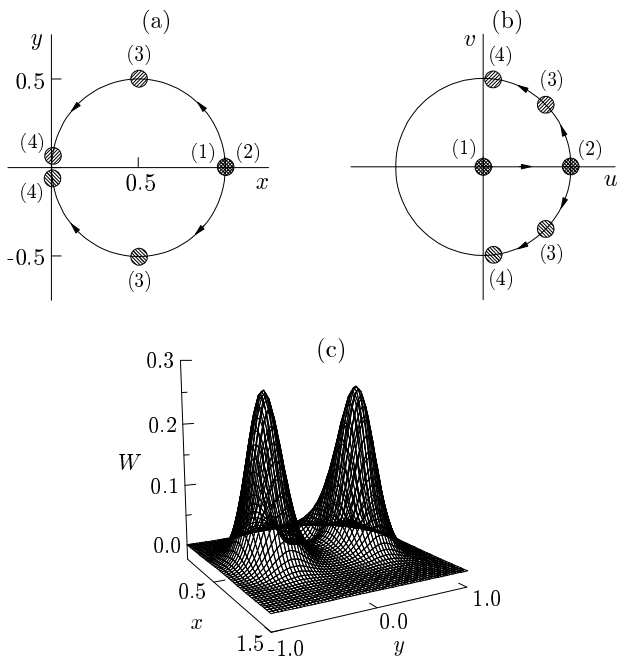}
\caption{Behavior of the semiclassical steady-state amplitudes $\alpha
/\bar\Omega$ (a) and $\beta$ (b) [Eq.~(\ref{eqn:amplitudes})] with
increasing $\bar\Gamma_a$; the labeled points are (1)~$\bar\Gamma_a=0$,
(2)~$\bar\Gamma_a=1/2\bar\Omega$, (3)~$\bar\Gamma_a=1/\sqrt2\bar\Omega$,
and (4)~$\bar\Gamma_a\gg1$. (c) Steady-state Wigner function for
$\bar\Gamma_a=1/\sqrt2\bar\Omega$ (maximum peak separation) for
$\bar\Omega=3$.} 
\label{fig:fig2}
\end{figure}

Alsing and Carmichael \cite{Alsing91} performed a semiclassical analysis
of Eq.~(\ref{eqn:master2}). They found that as a function of the 
parameter $\xi\equiv(2\bar\Gamma_a\bar\Omega)^{-1}$, the composite system
exhibits a symmetry breaking transition at $\xi=1$ (spontaneous
dressed state polarization). If $\alpha=\bar\Omega(x+iy)$ and $\beta=u+iv$
denote complex amplitudes of the cavity field and atomic polarization, in
steady state
\begin{equation}
\begin{matrix}
\qquad\alpha/\bar\Omega=1,\quad\beta=1/\xi,\hfill
&\hfill\qquad\mkern-3mu(\xi\geq1),\cr
\noalign{\vskip2pt}
\qquad\alpha_\pm/\bar\Omega=\xi\beta_\pm=\xi\left[\xi\pm i\sqrt{1-\xi^2}
\right],
&\hfill\qquad\mkern-3mu(\xi\leq1).
\end{matrix}
\label{eqn:amplitudes}
\end{equation}
Figures~\ref{fig:fig2}(a) and (b) illustrate the behavior of the steady-state
amplitudes as a function of $\bar\Gamma_a$. For $\bar\Gamma_a>1/2\bar\Omega$
($\xi<1$), there are two permissible values for the amplitude of $A$, each
correlated with an amplitude for $B$. With increasing $\bar\Gamma_a$, the
amplitudes of $A$ separate and move in opposite directions on a circle of
radius $\bar\Omega/2$; they reach maximum separation at $\bar\Gamma_a=1/\sqrt2
\bar\Omega$ and approach one another on the opposite side of the circle for
$\bar\Gamma_a\gg1$. The correlated amplitudes of $B$ move on a semicircle
as shown. If $\bar\Omega\gg1$, the phase-space separation across the
circle is large compared with the vacuum-state uncertainty of oscillator
$A$ [point (3) in Fig.~\ref{fig:fig2}(a)]. In such a case, the Wigner
function for the reduced quantum-mechanical steady state $\tilde\rho_A
={\rm tr}_B(\tilde\rho)$ is double peaked, as shown in Fig.~\ref{fig:fig2}(c).

\begin{figure}[t]
\includegraphics*[width=3.4in,keepaspectratio=true]{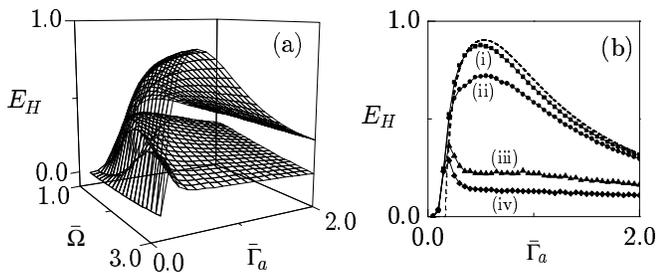}
\vskip-0.2cm 
\caption{Entanglement in homodyne detection: (a) as a function of
$\bar\Gamma_a$ and $\bar\Omega$ for $\theta=0$ (upper surface) and
$\theta=\pi/2$ (lower surface); (b) for fixed $\bar\Omega=3$ and
(i)~$\theta=0$, (ii)~$\theta=\pi/40$, (iii)~$\theta=\pi/10$, and
(iv)~$\theta=\pi/2$, the dashed line is the entanglement in ensemble
(\ref{eqn:ensemble2}). All results are for $\bar\Gamma_b=0$.} 
\label{fig:fig3}
\end{figure}

The full quantum-mechanical steady state $\tilde\rho$ is approximated,
for $\bar\Omega\gg1$, by an equally weighted mixture of the states
$|\alpha_+\rangle|\beta_+\rangle$ and $|\alpha_-\rangle|\beta_-\rangle$,
where $|\alpha_\pm\rangle$ are coherent states of oscillator $A$ and
the states of $B$ are $|\beta_\pm\rangle=\left(\sqrt{\beta_\pm}\,|e\rangle
+\sqrt{\beta_\pm^*}\,|g\rangle\right)/\sqrt2\,$, where $|e(g)\rangle$
denotes the excited (ground) state of the atom.
Two compatible pure state ensembles are 
\begin{equation}
|\psi_{\rm REC}\rangle=\Theta_{\rm REC}|\alpha_+\rangle|\beta_+\rangle
+(1-\Theta_{\rm REC})|\alpha_-\rangle|\beta_-\rangle,
\label{eqn:ensemble1}
\end{equation}
with $\Theta_{\rm REC}=0,1$ a dichotomous random variable (probabilities
$p_0=p_1=1/2$), and 
\begin{equation}
|\psi_{\rm REC}\rangle=N(e^{i\Phi_{\rm REC}}|\alpha_+\rangle|\beta_+
\rangle+e^{-i\Phi_{\rm REC}}|\alpha_-\rangle|\beta_-\rangle),
\label{eqn:ensemble2}
\end{equation}
with $\Phi_{\rm REC}$ uniformly distributed between 0 and $\pi$. For
ensemble (\ref{eqn:ensemble1}) the entanglement $E_U$ is zero; therefore
the approximate $\tilde\rho$ is separable. Nevertheless, the von Neumann
entropy of state (\ref{eqn:ensemble2}) is nonzero. We find it may be
computed from Eq.~(\ref{eqn:entropy}) by replacing $\lambda$ with
\begin{equation}
\lambda_\pm(\Phi_{\rm REC})=\frac12\pm\frac12\frac{\sqrt{A^2(\Phi_{\rm REC})
+B^2(\Phi_{\rm REC})}}{B(\Phi_{\rm REC})},
\label{eqn:analytical_approximation}
\end{equation}
where
\begin{equation}
A(\Phi_{\rm REC})=\sqrt{1-\xi^2}e^{-2\bar\Omega^2y^2}\sin[2(\bar\Omega^2xy
+\Phi_{\rm REC})],
\label{eqn:A}
\end{equation}
\begin{equation}
B(\Phi_{\rm REC})=\xi+e^{-2\bar\Omega^2y^2}\cos[2(\bar\Omega^2xy
+\Phi_{\rm REC})].
\label{eqn:B}
\end{equation}
After averaging over records a nonzero entanglement is obtained.
It is plotted as the dashed curve in Fig.~\ref{fig:fig3}(b), where
the behavior as a function of $\bar\Gamma_a$ resembles that
in Fig.~\ref{fig:fig1}(d), except Fig.~\ref{fig:fig3}(b) shows a
higher peak,
approaching unity for $\bar\Omega\gg1$. Clearly there is a vast
difference in the entanglement computed for ensembles (\ref{eqn:ensemble1})
($E_U=0$) and (\ref{eqn:ensemble2}) ($E_U\approx1$).

The difference has a physical origin; it arises from the way in
which information is read from the environment. Ensemble
(\ref{eqn:ensemble1}) yields $E_U=0$ because it assumes that this
information---encoded in $\Theta_{\rm REC}$---is able to
distinguish between $|\alpha_+\rangle$ and $|\alpha_-\rangle$. Ensemble
(\ref{eqn:ensemble2}) assumes the opposite---that the record read in the
environment cannot distinguish between $|\alpha_+\rangle$ and
$|\alpha_-\rangle$. Direct detection of scattered photons cannot
distinguish these states, for example, because $|\alpha_+|^2=|\alpha_-|^2$.
Neither can measuring the amplitude ${\rm Re}(\alpha_+)={\rm Re}
(\alpha_-)$. Measuring ${\rm Im}(\alpha_+)=-{\rm Im}(\alpha_-)$, on
the other hand, can.

More generally,
let us define the quadrature amplitude $\hat X_\theta=\left(e^{i\theta}
\hat a^{\dag}+e^{-i\theta}\hat a\right)/2$ and consider the
 homodyne detection  unraveling of $\tilde\rho$ (in the strong local
oscillator limit; $U=H$) corresponding to a $\theta$-quadrature measurement
of the light scattered into reservoir $R_A$. The pure state ensemble is
generated in this case by the stochastic Schr\"odinger equation~\cite{Carmichael93}
\begin{eqnarray}
d|\bar\psi_{\rm REC}\rangle=\left(\frac1{i\hbar}\tilde H_{\rm eff}d\bar t
+e^{-i\theta}\sqrt{2\bar\Gamma_a}\,\hat adq_\theta\right)\!|\bar\psi_{\rm
REC}\rangle,
\label{eqn:homodyne}
\end{eqnarray} 
where $\tilde H_{\rm eff}\equiv\tilde H_{AB}+\tilde H_{\rm ext}^B-\bar
\Gamma_a\hat a^\dagger\hat a$, and
\begin{equation}
dq_\theta=\sqrt{2\bar\Gamma_a}\langle e^{i\theta}\hat a
+e^{-i\theta}\hat a^\dagger\rangle_{\rm REC}d\bar t+d\bar W
\end{equation}
is the record of charge deposited in the detector; $d\bar W$ is a real Wiener
increment. Results for $E_H$ 
are displayed in Fig.~\ref{fig:fig3}. In Fig.~\ref{fig:fig3}(a) we plot
the maximum and minimum values of $E_H$ obtained with $\theta=0$
and $\pi/2$, respectively, as a function of $\bar\Omega$ and $\bar\Gamma_a$.
The minimum provides an upper bound on the EOF. It is not zero as
in ensemble (\ref{eqn:ensemble1}) because $\bar\Omega$ is finite.
In Fig.~(\ref{fig:fig3})(b), the maximum $E_H$ is
well-approximated by Eqs.~(\ref{eqn:analytical_approximation})-(\ref{eqn:B})
even for $\bar\Omega=3$. The optimal entanglement occurs along a curve
with $\bar\Omega$ inversely proportional to $\bar\Gamma_a$, which is what
one would expect from the definition of points (2) and (3) in Fig.~\ref{fig:fig2}.
Note, however, that with $\bar\Omega$ fixed, the peak, as a function of
$\bar\Gamma_a$, does not occur where $|\alpha_+\rangle$ and $|\alpha_-\rangle$
are maximally separated in phase space. This is explained by $|\beta_+\rangle$
and $|\beta_-\rangle$, which are not orthogonal at maximum separation, but
only approach orthogonality as $\bar\Gamma_a\to\infty$.

Returning to our general theme,
existing measures of entanglement view a mixed state $\rho$ as a fundamental
object and aim to associate a unique number with $\rho$. For open composite systems
the mixing arises from a well-defined process. In this case, more can be said
about entanglement by considering how
the system acts upon its environment---nonlocally in the case of emitted
light. Entanglement, as with the correlations it describes, becomes a
contextual notion. In this Letter we have shown how quantum trajectory theory
can quantify such entanglement, capturing the context-dependence in
its different unravelings of $\rho$. 

Many questions are left open concerning observable ramifications
of $E_U$ for different unravelings $U$. When the environment is monitored
as a means of conditional state preparation \cite{Bose99}, physical
implications are clear. More generally, do the records themselves
contain indications that the system generating the measured
outputs is described by an entangled state? The sensitivity of $E_U$
to minor changes in record making (imperfect detection)
is also an interesting topic, with implications for decoherence theory and
the classical limit. These and other issues are left for
future investigation.

This work was supported by the NSF under Grant No.\ PHY-0099576 and
by the Marsden Fund of the RSNZ.

\end{document}